\documentclass[useAMS,usenatbib,usegraphicx]{mn2e}

\usepackage{graphicx}
\usepackage[latin1]{inputenc}
\usepackage{color}
\usepackage{times}
\usepackage{natbib}
\usepackage{setspace}
\newif\ifAMStwofonts
\AMStwofontstrue
\definecolor{red}{rgb}{1,0.,0.}

\newcommand{\muvlim}{$M^{\rm lim}_{145}$}
\newcommand{\fesc}{$f_{\rm esc}$}
\newcommand{\hfesc}{$f^{\rm faint}_{\rm esc}$}
\newcommand{\tfesc}{$f^{\rm agn}_{\rm esc}$}

\def\lesssim{\lower.5ex\hbox{$\; \buildrel < \over \sim \;$}}
\def\gtrsim{\lower.5ex\hbox{$\; \buildrel > \over \sim \;$}}
\title[Cosmic Ionizing Background and \fesc] {On the Evolution of the Cosmic
  Ionizing Background.}

\author[Fontanot et al.]{
  \parbox[t]{\textwidth}{Fabio Fontanot$^{1,2,3}$\thanks{E-mail:
      fontanot@oats.inaf.it}, Stefano Cristiani$^{3,4}$, Christoph
    Pfrommer$^{1}$, Guido Cupani$^3$, \\ Eros Vanzella$^5$}
    \vspace*{8pt}\\
    $^1$ HITS-Heidelberger Institut f\"ur Theoretische Studien, Schloss-Wolfsbrunnenweg 35, 69118 Heidelberg, Germany\\
    $^2$ Institut f\"ur Theoretische Physik, Philosophenweg, 16, 69120, Heidelberg, Germany \\
    $^3$ INAF - Astronomical Observatory of Trieste, via G.B. Tiepolo 11, I-34143 Trieste, Italy \\
    $^4$ INFN - National Institute for Nuclear Physics, Via Valerio 2, I-34127 Trieste, Italy \\
    $^5$ INAF - Astronomical Observatory of Bologna, via E. Ranzani 1, I-40127, Bologna, Italy\\
}

\begin{document}
\date{Accepted ... Received ...}

\maketitle

\begin{abstract} 
We study the observed cosmic ionizing background as a constraint on
the nature of the sources responsible for the reionization of the
Universe. In earlier work, we showed that extrapolations of the
Ultra-Violet Luminosity Function (LF) of Lyman Break Galaxies (LBGs)
at fixed Lyman continuum photon escape fraction (\fesc) are not able
to reproduce the redshift evolution of this background. Here, we
employ extrapolations of the high-z LFs to describe the contribution
of LBGs to the ionizing photon rate, taking into account the smoothing
of the baryonic perturbations, due to the background itself (i.e. the
filtering mass), as well as a possible sharp increase of \fesc~in
dwarf galaxies. Under the hypothesis of a dominant contribution of
LBGs to cosmic reionization, our results suggest that sources fainter
than the current observational limits should be characterised by
\fesc~values of the order of $\sim 0.1-0.3$ (larger than the current
estimates for bright galaxies) to account for a $z \gtrsim 6$
reionization and the measured evolution of cosmic ionizing background,
{\it at the same time}. The contribution to the background from
quasars turns out to be relevant at $z \lesssim 3$. Overall, our
results support the case for dedicated observations of faint galaxies
in the rest-frame UV, in order to better determine their physical
properties. Observed \fesc~values outside our proposed range bear
relevant consequences on the nature of the astrophysical sources
responsible for cosmic reionization and/or its buildup process.
\end{abstract}

\section{Introduction}\label{sec:intro}

The epoch of cosmic reionization is a crucial signpost to understand
the formation of the first cosmic structures, the thermal history of
the inter-galactic medium (IGM) and the interaction between energetic
photons and the different gas phases (which are responsible for the
onset of processes like gas cooling and star formation). A number of
astrophysical evidences, ranging from the analysis of the cosmic
microwave background \citep{WMAP7, Planck_cosmpar}, to the
Gunn-Peterson test applied to the Lyman $\alpha$ forest
\citep{McGreer11}, to the fraction of Lyman $\alpha$ emitters among
$z\sim7$ Lyman Break Galaxies (LBGs, \citealt{Pentericci11}) broadly
constrain the epoch of Hydrogen reionization at $6<z<12$ (with a peak
probability at $z\sim10$ if instantaneous reionization is assumed).

On the other hand, the nature of the astrophysical sources responsible
for the reionization is still a matter of considerable debate. Star
forming galaxies and active galactic nuclei (AGNs) have been usually
suggested as the main contributors to reionization and their role has
been the subject of extensive research \citep{Haiman98,
  SchirberBullock03, Robertson10, Bouwens11b}. Nonetheless, the global
details of the process are still elusive and one of the major puzzles
is the evidence that {\it given the properties of observed galaxies
  and AGNs} none of the known astrophysical populations is able of
account {\it alone} for the whole ionizing photon budget required to
complete reionization at $z>6$, thus not excluding a relevant
contribution of more exotic sources \citep{Scott91, Madau04,
  Pierpaoli04, Dopita11}. We recently reviewed the combined
contribution of both galaxies and AGNs to reionization
\citep[hereafter FCV12]{Fontanot12b}: we showed that AGNs alone cannot
be responsible for cosmic reionization, but they can indeed provide a
relevant, although sub-dominant contribution to the photon budget,
thus alleviating the constraints for the LBG contribution. In fact, we
also showed that the required ionizing photon density can be produced
by the LBG population, only if their {\it escape fractions} (\fesc,
i.e. the fraction of ionizing photons, which are able to escape the
local environment and ionize the IGM) is of the order of 0.2 and/or a
substantial contribution is due to faint galaxies mostly beyond the
reach of present technology \citep[see also][]{Bouwens11b, Kuhlen12,
  Alvarez12, Robertson13}.

In particular, \fesc~is a key parameter for the determination of the
relative contribution of different galaxy populations to reionization,
but it is very poorly constrained by direct observations. Due to their
hard spectra, AGNs are supposed to have large escape fractions,
possibly reaching unity in the most luminous QSOs
($M_B<-23$). However, it is not well established yet if such a
conclusion holds also for fainter sources (i.e. AGNs), given that the
fraction of obscured (Type-II) sources increases with magnitude
\citep[see e.g.,][]{Simpson05}. On the other hand, galaxies are
characterised by softer spectra blue-ward of Ly$\alpha$ and their
\fesc~is expected to be lower than for AGNs, given the presence of
cold gas and dust, which absorb most of the Lyman continuum emission
\citep[see e.g.,][]{Haehnelt01}. Direct detection of Lyman continuum
photons at $z<1.5$ has been elusive so far
\citep[\fesc$<0.01$][]{Leitherer95, Cowie09}, while estimates at $z
\sim 3-4$ cover a wide range of results, from low ($f_{\rm esc} <
0.05$, \citealt{Vanzella10b}), to relatively high values ($f_{\rm esc}
\gtrsim 0.2$, \citealt{Iwata09,Nestor13}, but see
\citealt{Vanzella12a}). From the theoretical side, numerical
simulations provide contrasting suggestions, reporting opposite
(i.e. negative and positive) dependencies on the environment and a
relevant object-to-object scatter \citep{Gnedin08, Yajima11}.

A strong constraint on the evolution of ionizing sources comes from
the measure of the intensity of the ionizing background: the observed
$2<z<6$ photoionization rate per unit comoving volume is only a few
times larger than the required minimum value for the reionization of
the Universe \citep{MHR99, Pawlik09}, thus narrowing the allowed
parameter space for \fesc~and the limiting magnitude at 145 ${\rm nm}$
of the luminosity function (\muvlim) of the contributing sources. In
FCV12 we showed that the observed redshift evolution of the
astrophysical sources translates into an evolution of the predicted
photoionization rates contrasting with the observational data, thus
exacerbating the difficulties in reproducing the background {\it and}
achieving reionization at $z>7$. In particular, {\it at fixed}
\fesc~the background constraint implies strong limits for \muvlim~and
vice-versa, thus favouring models which assume a redshift (and/or
luminosity) evolution for \fesc, \muvlim~or both.

Indeed, several authors \citep[see e.g.,][]{Ciardi12} proposed a
variable \fesc~as a function of redshift and/or luminosity as a viable
solution to avoid a relevant contribution to reionization from faint
sources (which also implies an extended reionization epoch,
\citealt{Kuhlen12}). A more extreme scenario requires reionization to
be dominated at $z\gtrsim9$ by ultra-faint galaxies: these sources
should be characterised by higher \fesc~values (close to unity) and
live in $10^8-10^9 M_\odot$ host dark matter (DM) haloes. This implies
that these sources would not have formed at $z \lesssim 6$ in ionized
regions (therefore not contributing to the background), due to
suppression by heating from the UV background \citep[see
  e.g.][]{BabulRees92, Gnedin00, BarkanaLoeb02, Somerville02,
  CiardiFerrara05, Okamoto08}. Such an important hypothesis is clearly
difficult to test observationally due to the increasing complexity of
a direct measurement of \fesc~for faint objects at high redshift
\citep{Vanzella12b}. In this paper, we employ toy models, based on the
observed evolution of AGN and LBG LFs, in order to build a plausible
scenario for the reionization and the production and evolution of the
UV background, as well as to tailor observational tests to be
performed on the faintest galaxies accessible.

This paper is organised as follows. In Section~\ref{sec:models}, we
present the formalism to estimate the ionizing background from
observed LFs. This will be used in Section~\ref{sec:results} to
provide constraints on the expected \fesc~value for faint sources and
we discuss our conclusions in Section~\ref{sec:concl}. Throughout this
paper we assume that quasi-stellar objects (QSOs) represent the
luminous sub-population of the homogeneous AGN population, and that
LBGs are a good tracer of the overall galactic population.

\section{Synthesis of the Ionizing Background}\label{sec:models}

We consider two different estimates of the ionizing background, in
order to constrain the expected physical properties of the sources
responsible for cosmic reionization. We thus compare the
observationally determined photoionization rates as a function of
redshift, with the prediction of theoretical models for the synthesis
of the background, based on the redshift evolution of the luminosity
functions (LFs) of different astrophysical sources (i.e. LBGs and
AGNs). It is worth stressing that the two estimates of the background
are completely independent. On the one hand, the direct determinations
of the photoionization rates are based on the observed
Ly$\alpha$-forest effective opacity \citep{Bolton07, Becker07,
  FaucherGiguere08, WyitheBolton11, BeckerBolton13} and on the QSO
proximity effect \citep{Giallongo96, Dodorico08, Calverley11}. In most
cases, the hydrogen photoionization rates are derived from the mean
opacity of the IGM to Ly$\alpha$ photons (quantified in terms of an
effective optical depth), via a determination of the temperature of
the IGM. The large scatter between the different determinations of the
ionizing background is mainly due to the different assumptions for the
IGM temperature, Ly$\alpha$ and ionizing opacities among the different
authors \citep[see e.g.,][and discussion herein]{BeckerBolton13}: it
is then indicative of our limited knowledge of the properties of the
IGM at different redshifts. On the other hand, our theoretical
determination of the expected ionizing background does not require any
assumption on the detailed properties of the IGM (like its temperature
and density), but it is mainly based on our understanding of galaxy
evolution; therefore its main uncertainties are tightly linked to our
limited knowledge of the statistical properties of the galaxy
populations and how UV photons are able to escape from them (\fesc,
\muvlim~and the faint-end slope of the LF). In the following, we will
take advantage of these independent estimates to explore the
theoretical parameter space and put the tightest possible constraints
on \fesc.

\subsection{AGN/LBG contribution to the ionizing background}
\begin{figure}
  \centerline{\includegraphics[width=9cm]{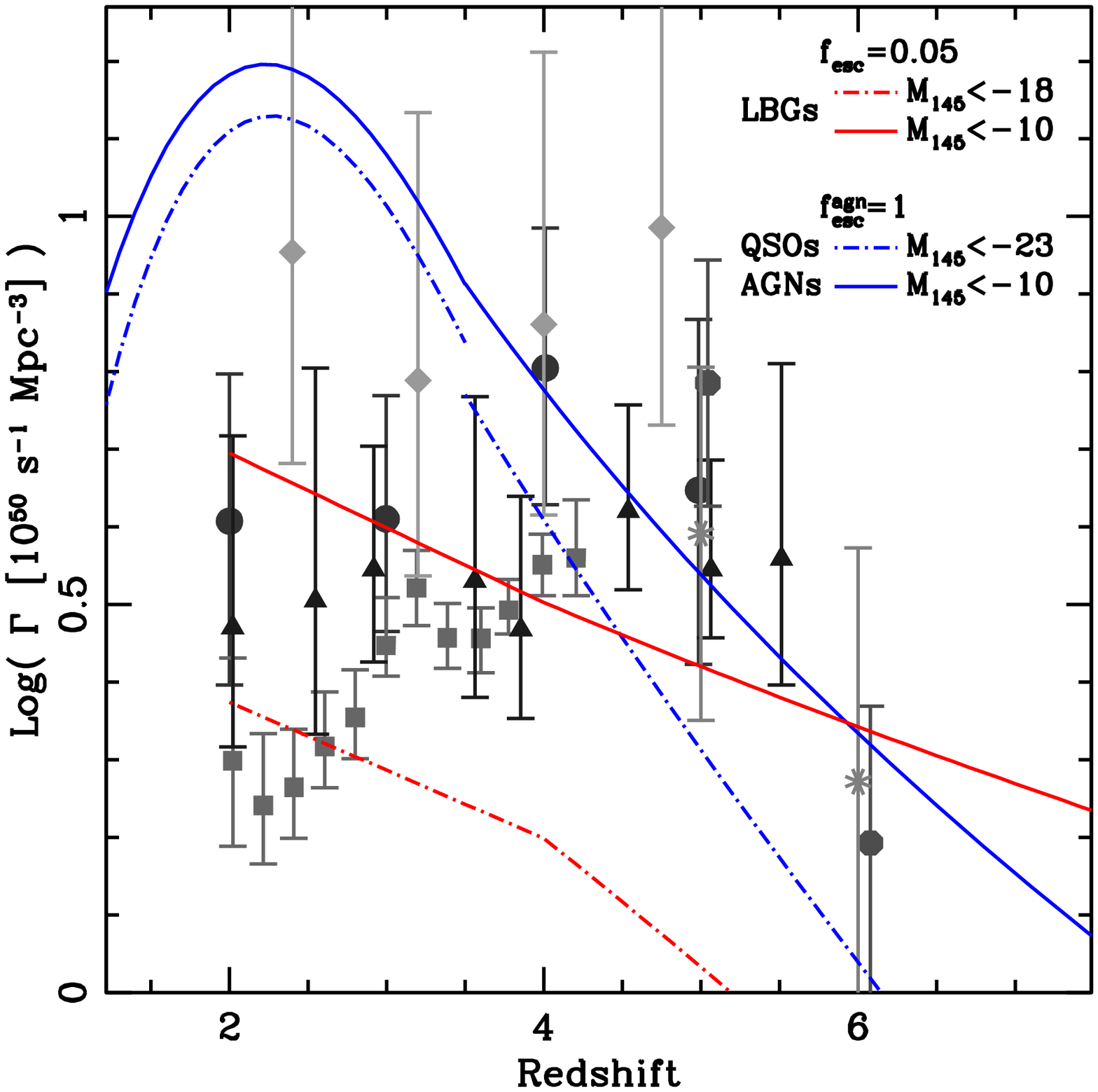}}
  \caption{Contribution of different galaxy populations to the
    ionizing background. Observed data from
    \citet[][circles]{Bolton07}, \citet[][triangles]{Becker07},
    \citet[][squares]{FaucherGiguere08},
    \citet[][pentagons]{Calverley11},
    \citet[][asterisks]{WyitheBolton11},
    \citet[][diamonds]{BeckerBolton13}. The blue lines refer to
    the expected contribution of AGNs/QSOs as estimated from the
    \citet{Richards05} and \citet{Fiore11} LFs assuming \tfesc$=1$,
    with a discontinuity between the two at $z=3.5$. The red lines
    correspond to the contribution from LBGs as estimated from the
    \citet{ReddySteidel09} and \citet{Bouwens11a} LFs assuming
    \fesc$=0.05$. Dashed and solid lines correspond to the expected
    contribution of sources brighter than the limit indicated in the
    magnitude legend. }\label{fig:obs}
\end{figure}

As in FCV12, we estimate the expected contribution of different galaxy
populations to the ionizing background, by considering a reference set
of LFs at different redshifts. In particular, we focus on the
determination of the bolometric $z<6$ AGN-LF from \citet{Hopkins07b},
on the X-ray selected $3<z<7$ AGN-LF by \citet{Fiore11}, on the
$2<z<3$ and $3<z<8$ UV-LFs of LBGs by \citet{ReddySteidel09} and
\citet{Bouwens11a}, respectively. As in FCV12 we first compute the
monochromatic luminosity density $\rho_\nu$ from a given luminosity
function, and we use it to estimate both the LBG rate of emitted
ionizing photons per unit comoving volume as a function of redshift,

\begin{equation}\label{gamma_lbg}
\Gamma_{\rm LBG}(z) = f_{\rm esc} \kappa  \rho_{\rm UV}(z) / \zeta \, ,
\end{equation}

\noindent
and the corresponding AGN rate,

\begin{equation}\label{gamma_agn}
\Gamma_{\rm AGN}(z) = f^{\rm agn}_{\rm esc} \int_{\nu_H}^{\nu_{\rm up}}
\frac{\sigma_\nu \rho_\nu(z)}{h_p \nu} \, d \nu \, .
\end{equation}

\noindent
In equation~(\ref{gamma_lbg}), we define the conversion
factor\footnote{In equation~\ref{gamma_lbg}, we implicitly assume, as
  in FCV12, a first conversion from $\rho_{\rm UV}$ to the star
  formation rate density $\rho_{\rm SFR}$, which is then used to
  estimate $\Gamma_{\rm LBG}$. The conversion factors $\zeta$ and
  $\kappa$ critically depend (up to roughly $50\%$) on the assumptions
  on the properties of the underlying stellar population, like its
  Initial Mass Function, metallicity and dust content. An alternative
  approach has been proposed by, e.g., \citet{Kuhlen12}, which is
  based on an assumption of the typical galaxy spectral energy
  distribution and a single conversion from $\rho_{\rm UV}$ to the
  photon ionizing luminosity. Here we note that the combination
  $\kappa / \zeta = 1.20 \times 10^{25} \, {\rm erg}^{-1} {\rm Hz}$ is
  in good agreement with both the fiducial parameter used in
  \citet{Kuhlen12} and the range of values for different choice of
  stellar population parameters discussed in \citet[][their
    Fig.~1]{Robertson13}.} (Salpeter IMF) $\zeta = 1.05 \times 10^{28}
[{\rm erg} \, {\rm s}^{-1} \, {\rm Hz}^{-1} \, {\rm M_\odot}^{-1} \,
  {\rm yr}]$ from UV luminosity density to the star formation rate
density and $\kappa=10^{53.1} {\rm s}^{-1}$ is the number of LyC
photons produced per unit star formation rate (see e.g.,
\citealt{Shull12}). In equation~\ref{gamma_agn}, $\nu_H$ is the
frequency at $912$ \AA, and $\nu_{\rm up} = 4 \nu_H = 12.8 \times
10^{15} {\rm Hz}$ (assuming a cutoff at 4 Ryd, since more energetic
photons are mainly absorbed on He II, \citealt{MHR99}). The absorbing
cross section for neutral hydrogen $\sigma_\nu$ is assumed to be unity
between $\nu_H$ and $\nu_{\rm up}$, and zero outside this range. In
practice, we consider in the calculations a QSO spectral continuum of
the form $f_{\nu} \propto \nu^{-1.76}$ blue-ward of the Ly$\alpha$
line, as suggested by \citet{Telfer02}, where $f_{\rm nu}$ represents
the monochromatic flux at the frequency ${\rm \nu}$. In both equations
\fesc~and \tfesc~represent the escape fraction for the LBG and AGN
population, respectively. It is worth stressing that most
observational works prefer to use a {\it relative} escape fraction,
i.e. the fraction of escaping LyC photons, relative to the fraction of
escaping non-ionizing ultraviolet photons. Since in this quantity dust
attenuation is already taken into account, it is commonly used to
convert the observed luminosity density at $1500$ \AA~into Lyman
continuum emission. For the sake of simplicity, in the following we
will neglect any dust correction in our computations, thus making the
two definitions equivalent in our context.

In Fig.~\ref{fig:obs} we compare our empirical predictions with a
collection of observational constraints (grey dots) obtained by
converting photoionization rates, into ionizing photons per unit
comoving volume, following the same procedure and assumptions as in
\citet{Kuhlen12}. The blue lines correspond to the predicted
$\Gamma_{\rm AGN}$, assuming \tfesc$=1$. Solid and dashed lines
correspond to different lower limits in the integration of the
luminosity function (\muvlim$<-10$ and \muvlim$<-23$ respectively, the
usual discriminator between AGNs and QSOs). Red lines\footnote{We
  smooth the redshift evolution of the binned LBG-LFs at $z<4$ by
  requiring a linear evolution of the break magnitude ($M_\star$) and
  normalization ($\phi_\star$) of the LF between the $z=2.5$ value as
  in \citet{ReddySteidel09} and the $z=4$ value as in
  \citet{Bouwens11a}, at fixed faint-end slope ($\alpha=-1.73$). At
  $z>4$ we use the evolutionary model of \citet{Bouwens11a}.} show the
corresponding predictions for $\Gamma_{\rm LBG}$, assuming
\fesc$=0.05$. Also in this case, solid and dashed lines correspond to
different lower limits in the integration of the luminosity function
(\muvlim$<-10$ and \muvlim$<-18$, respectively). In both cases, we
extrapolate the observed LFs beyond their magnitude limits, assuming
that their faint-end slope estimates are a good proxy for the
statistical properties of fainter galaxies. We discuss uncertainties
associated with this extrapolation later in this work.

As already suggested in FCV12 the redshift evolution of the
contribution of ionizing radiation coming from both the galaxy and AGN
populations evolves differently in comparison to the observed
evolution of the background ({\it at fixed} \fesc). The observed
background shows almost no evolution (or at most a mild increase) with
redshift in the range $2<z<6$, while simple models predict a strong
decrease of all background contributions with redshift: there is no
straightforward way to reconcile, at fixed \fesc, the predicted
redshift evolution of the ionizing background with the data by
changing the values of the main parameters in
equation~(\ref{gamma_agn}) and~(\ref{gamma_lbg}). In particular, we
notice that the high \tfesc~values required for a significant
contribution of AGNs to the ionizing background at $z>4$ correspond to
a large AGN contribution to the background at $z\sim2$, where QSOs are
expected to dominate the emission of ionizing photons. These large
ionizing rates are in agreement with recent estimates by
\citet{BeckerBolton13}, but systematically exceed earlier results
(i.e. observations by \citealt{Bolton07} and
\citealt{FaucherGiguere08}); even if most of the data are still
compatible with the model at a $\sim 2 \sigma$ level.

\subsection{Suppression Galaxy Formation through IGM heating}

A possible solution to the tension shown in Fig.~\ref{fig:obs} lies in
postulating an evolution of \fesc~with redshift and/or
luminosity. However, the theoretical scenario is uncertain: star
formation, feedback, metallicity and dust content, the geometry and
distribution of the ISM are all important ingredients that have to be
taken into account; moreover, an explicit treatment of radiative
transfer is required to follow the ionization history in
detail. Contrasting results have been presented in the literature,
suggesting that \fesc~can either increase \citep{Razoumov06,
  FerraraLoeb13} or decrease \citep{WoodLoeb00} with redshift; or that
it decreases with increasing \citep{RicottiShull00, Yajima11} or
decreasing \citep{Gnedin08, FerraraLoeb13} halo mass. In general, in
order to achieve reionization with LBGs, it is necessary to postulate
that \fesc~increases for dwarf galaxies and/or in smaller haloes, with
respect to brighter and more massive systems. In the following we test
a possible alternative scenario that is able to reconcile the present
constraints from the LFs with the observational determination of the
ionizing background. It is based on a redshift dependent integration
limit for the LBGs UV-LF.

In fact, the presence of a background UV flux provides photoheating to
the gas, which has strong effects on galaxy formation in small dark
matter (DM) haloes. This delays cooling and reduces the efficiency of
galaxy formation in these environments since pressure support prevents
gas in halos with a circular velocity lower than the sound speed of
ionized gas from collapsing: baryonic fluctuations on small scales
thus grow at a slower rate than the corresponding DM fluctuations. It
is possible to describe this mechanism in linear theory by introducing
a characteristic ``filtering'' scale $\lambda_F = 2 \pi / k_F$
\citep{GnedinHui98} for the smoothing of baryonic fluctuations, which
is related to the linear growth function $D_+$, the scale factor $a$,
the Hubble function $H$ by the following equation:

\begin{equation}\label{eq:filtering}
\frac{D_+(t)}{k^2_F(t)} = \int_0^t dt' a^2(t') \frac{\ddot{D}_+(t') + 2 H(t') 
\dot{D}_+(t')}{k_J^2(t')}  \int_{t'}^t \frac{dt''}{a^2(t'')} \, .
\end{equation}

\noindent 
where $k_J$ represents the Jeans wave number defined as:

\begin{equation}\label{eq:jeans}
k_J(a) = \left( \frac {4 \pi a^2 \, G \, \bar{\rho}(a)}{5 \, k \, T(a)
  \, / \, 3 \, \mu \, m_p} \right)^{1/2}
\end{equation}

\noindent
where $T$ denotes the temperature at the mean total mass density of
the Universe $\bar{\rho}$, $k$ is the Boltzmann constant, $\mu$ the
mean molecular weight, $m_p$ the proton mass and the denominator
corresponds to the linear sound speed. It is thus straightforward to
define the {\it filtering mass} ($m_F$) as the mass enclosed in a
sphere with comoving radius $2 \pi a / k_F$ and it corresponds to the
mass of the halo which would lose half of the baryons, compared to the
universal baryon fraction \citep{Gnedin00}. The filtering mass is thus
a convenient indicator of the mass scale at which galaxy formation
becomes inefficient: therefore we assume that it should correspond to
the typical DM halo mass that hosts the fainter galaxies whose
statistical properties are still consistent with the observed
UV-LFs. We thus assume that for smaller DM haloes galaxy formation
becomes so inefficient that those haloes do not contribute to the
ionizing background.

\begin{figure}
  \centerline{\includegraphics[width=9cm]{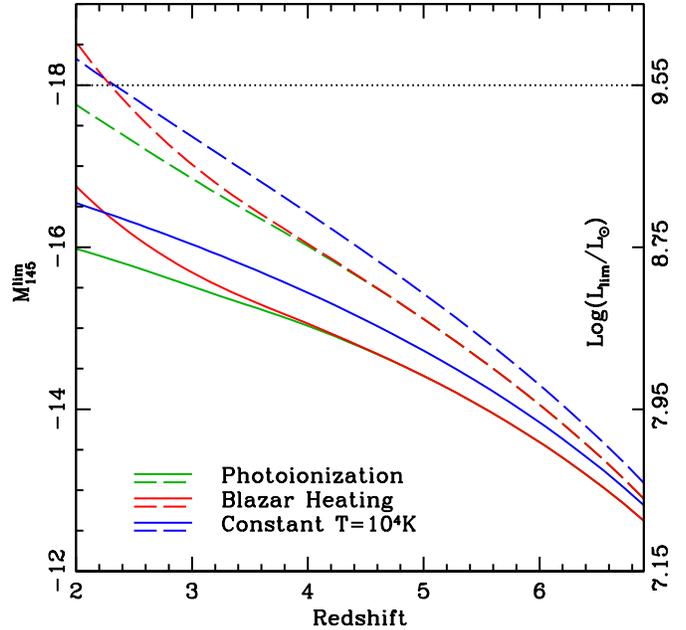}}
  \caption{Limiting magnitudes (luminosities) corresponding to the
    different determination of the filtering (dashed lines) and
    characteristic (solid) mass, according to the conditional UV-LF of
    \citet[][see text for more details]{Trenti10}. Green, red and blue
    lines refer to the different thermal histories assumed in solving
    equation~(\ref{eq:filtering}) as indicated in the legend. The
    horizontal dotted line represents the current limit for faint LBG
    searches \citep{Bouwens11a}.}\label{fig:m145lim}
\end{figure}

The definition of $m_F$ depends on the details of the assumed thermal
history and in the following we will discuss results relative to three
different choices, covering a reasonable range of
possibilities. Recently, there has been another non-gravitational
heating mechanism suggested in addition to photoheating. TeV photons
emitted by luminous hard blazars pair produce and annihilate on the
extragalactic background light, which is emitted by galaxies and
QSOs. Powerful plasma instabilities appear to be able to dissipate the
kinetic energy of the electron-positron pairs at a rate that is faster
than the inverse Compton cooling rate \citep{Broderick12,
  Schlickeiser12, Schlickeiser13}. This leads to a temperature-density
relation that is partially inverted toward the low-density voids
\citep{Chang12}, which is in agreement with all statistics on the
Lyman-alpha forest \citep{Puchwein12}. Additionally, this implies a
dramatic increase in the entropy of the IGM following He {\sc ii}
reionization at $z \sim 3.5$ and increases the filtering mass
\citep{Pfrommer12}. Following \citet{Pfrommer12}, we consider a pure
photoheating (green lines in Fig.~\ref{fig:m145lim}
and~\ref{fig:mainO}) and a ``standard'' blazar heating (blue lines)
model with sudden reionization for H {\sc i} (at $z = 10$) and He {\sc
  ii} (at $z = 3.5$). Thermal and entropy histories for these models
have been extensively discussed in \citet[][see their
  Figure~1]{Pfrommer12}. Finally we consider a model with constant
($10^4 {\rm K}$) temperature (red lines).

The filtering mass obtained through equation~(\ref{eq:filtering}) is
an estimate of the relevant mass scale in linear perturbation
theory. However, the exact mass scale below which galaxy formation
becomes inefficient depends also on the details of non-linear
structure formation, which can be followed accurately by means of
cosmological $N$-body simulations \citep[see e.g][]{Hoeft06,
  Okamoto08}. These results predict a {\it characteristic mass}
($m_C$), defined as the halo mass at which haloes have lost (on
average) half of their baryons, roughly one order of magnitude smaller
than the filtering mass estimate at $z=0$. It is worth stressing that,
according to the previous definition, $m_C$ does not correspond to the
mass scale at which the production of ionizing photons ceases (as in
the filtering mass approach). In the following we will assume that
$m_C$ marks the mass scale at which the faint-end of the LBG-LF drops
and the contribution of fainter galaxies to reionization becomes
negligible; we thus assume an effect similar to $m_F$. In order to
capture the (redshift dependent) relation between $m_C$ and $m_F$, we
follow \citet{Maccio10} and we introduce a correction factor that was
calibrated to results by \citet[][from their Fig.~B1]{Okamoto08}:

\begin{equation}
\frac{m_C}{m_F}(z) = \frac{(1+z)^{1.1}}{11.8} \, .
\end{equation}

We then consider the problem of estimating the UV luminosity $L_{\rm
  lim}$ of the typical galaxy hosted in a halo of mass $m_F$ or
$m_C$. To this aim we use the improved conditional luminosity function
proposed by \citet{Trenti10}, which we describe as:

\begin{equation}\label{eq:trenti}
\frac{L_{\rm lim}}{10^{8.87} L_\odot} = \left( \frac{m_{\rm h}}{10^{10} M_\odot}
\right)^{1.3} \, .
\end{equation}

\noindent
The resulting limiting magnitudes corresponding to our different
choices of filtering/characteristic mass are shown in
Fig.~\ref{fig:m145lim}. It is important to keep in mind that this
improved conditional luminosity function is defined for $4 \lesssim z
\lesssim 6$ and we are extrapolating it to lower redshifts.  Moving
from the filtering to the characteristic mass definition implies a
magnitude difference of $\sim 2 \, {\rm mag}$, while the difference
between the difference thermal histories are smaller. In particular,
blazar heating differs from photoionization heating only for $z
\lesssim 3$ and accounting for almost a magnitude difference at $z
\sim2$. These results show that the different definitions of the
limiting DM halo masses are more relevant for the estimate of
$\rho_{UV}$, than the details in the assumed thermal history.

\section{Results \& Discussion}\label{sec:results}

For each definition and redshift evolution of the
filtering/characteristic mass, we compute the ionizing background, by
integrating the UV-LF up to the corresponding $L_{\rm lim}$. Since the
resulting \muvlim~at $z < 7$ is brighter than $-10$, it is not
possible to achieve reionization at $z \sim 6$ and reproduce the
ionizing background with a constant \fesc$=0.05$ over the whole
magnitude range (see discussion in FCV12). We thus explore the effect
of an increase of \fesc~for the fainter galaxies. For the purpose of
the present work, we assume a sharp increase in \fesc~(\hfesc) for all
galaxies with $M_{\rm break} > -18$ (the present limit of faint LBG
searches according to \citealt{Bouwens11a}), while keeping
\fesc$=0.05$ for brighter galaxies:

\begin{equation}
f_{\rm esc} = \left\{
\begin{array}{ll}
0.05 & \textrm{if $M_{145} < M_{\rm break}$} \\
f^{\rm faint}_{\rm esc} & \textrm{if $M_{145} > M_{\rm break}$} \\
\end{array}
\right.
\label{eq:hfesc}
\end{equation}

\begin{figure}
  \centerline{ \includegraphics[width=9cm]{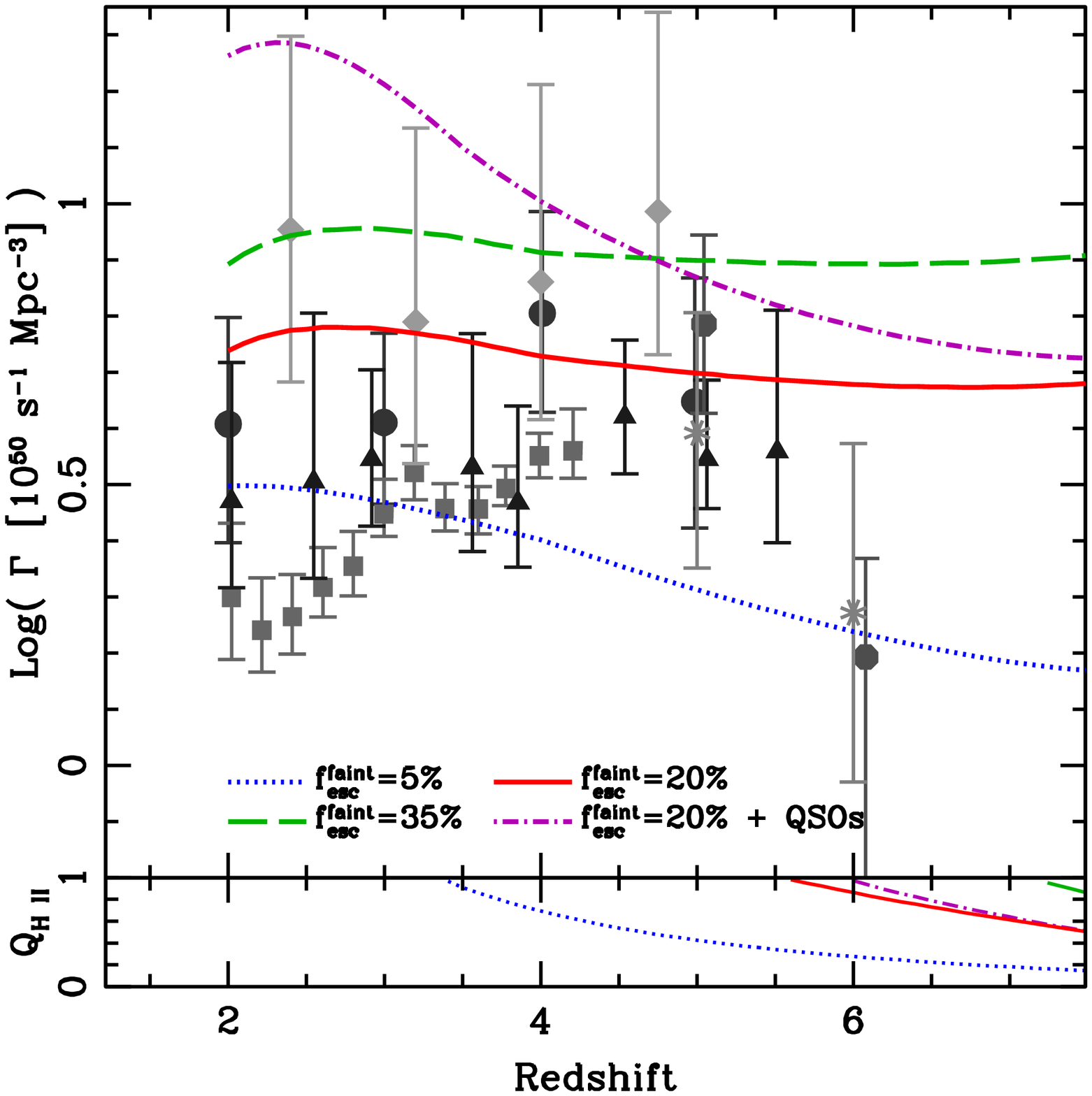} } 
  \caption{{\it Main panel:} synthesis of the ionizing
    background. Datapoints as in Fig.\ref{fig:obs}. Dotted, solid and
    dashed lines represent the LBG contribution obtained by assuming,
    respectively, \hfesc$=0.05$, $0.2$ and $0.35$. The dot-dashed
    purple line represents the total contribution of LBGs
    (\hfesc$=0.2$) and QSOs (\muvlim$<-23$) to the ionizing
    background. For all lines we consider the blazar heating thermal
    history and characteristic mass case. {\it Bottom panel:} redshift
    evolution of the filling factor of H {\sc ii} regions for the same
    models.}\label{fig:fevo}
\end{figure}

\noindent
We want to study whether this hypothesis is able to reconcile the
shape of the evolution of the ionizing background with reionization at
$z \sim 6$ and infer the expected range of \fesc values in LBGs. We
propose to test such a prediction with dedicated observations (to be
performed with present and future facilities). In the following, we
define a reference model (i.e. a thermal history which accounts for
photo- and blazar heating for the case of the characteristic mass) and
discuss the dispersion of our predictions, against variations of the
main parameters of our approach. In Fig.~\ref{fig:fevo}, we then show
how our predictions scale with varying \hfesc: we consider
\hfesc$=0.05$ (the upper limit for $z\sim3$ estimates of bright
galaxies), \hfesc$=0.2$ (a value often advocated to achieve cosmic
reionization by LBGs only, see e.g \citealt{Bouwens11b}) and a higher
\hfesc$=0.35$ (to explore the limits of our approach). As a reference,
we compute the reionization epoch associated to our toy models by
considering in the bottom panel the filling factor of reionized
hydrogen ($Q_{\rm H II}$) solving the following differential equation
(as in \citealt{MHR99}):

\begin{equation}\label{eq:qh2}
\dot{Q}_{\rm H II} = \frac{\Gamma}{\bar{n}_H} - \frac{Q_{\rm H
    II}}{\bar{t}_{\rm rec}}
\end{equation}

\noindent
where $\bar{n}_H$ is the mean comoving density of hydrogen atoms and
the equation for $\bar{t}_{\rm rec}$ has been computed following
\citet{Kuhlen12}:

\begin{equation}\label{eq:trec}
\bar{t}_{\rm rec} [{\rm Gyr}] \simeq 0.93 \, \frac{3}{C(z)} \, \left(
\frac{T(a)}{2 \times 10^4 K} \right)^{0.7} \left( \frac{1+z}{7}
\right)^{-3} 
\end{equation}

\noindent
assuming for each toy model the appropriate thermal history and the
redshift dependent clumping factor $C(z)$ derived in
\citet{HaardtMadau12}:

\begin{equation}\label{eq:cfac}
C(z) = 1 + 43 \times z^{-1.71}
\end{equation}

\noindent
This gives typical values of the order unity, which is in line with
the other recent estimates of the clumping factor for gas with
overdensities below 100 \citep[see e.g.,][]{Pawlik09, Shull12}. Using
the above calculations it is also possible to estimate the Thomson
scattering optical depth to cosmic microwave background, $\tau$:

\begin{equation}
\tau = \int_0^\infty dz \frac{c (1+z)^2}{H(z)} \, \sigma_T \,
\bar{n}_{\rm H} \left( 1+\frac{\eta Y}{4(1-Y)} \right)
\end{equation}

\noindent
where $\sigma_T$ represents the Thomson cross section and $Y$ is the
fraction of mass in Helium, which we assume to be singly ionized
($\eta=1$) at $z>3.5$ and doubly ionized ($\eta=2$) at lower
redshift.

From our analysis, it is evident that both ingredients (i.e. a
redshift-dependent $L_{\rm lim}$ and \hfesc$>0.05$) are necessary to
recover the redshift evolution of the observed background together
with reionization at $z>6$ (which we assume to happen in our toy
models, when $Q_{\rm H II}$ reaches unity in the bottom panel).
Comparison with the data points clearly favour \hfesc$> 0.05$: the
lower bound is well constrained by the reionization epoch, while the
upper bound is relatively unconstrained, due to the systematic
uncertainties in the observed $\Gamma$; \hfesc$=0.20$ agrees
reasonably well with most of the points over a wide redshift
range. These conclusions are confirmed by the predicted $\tau$ values
for our toy models: they are quite low ($\tau=0.035$) for
\hfesc$=0.05$, and consistent\footnote{While a value of $\tau\sim0.06$
  is statistically consistent with CMB {\em Planck} data at the
  1-$\sigma$ confidence level, we caution that the inferred values for
  $\tau$ may also be an underestimate owing to the simplified
  assumption of a symmetric reionization history. An extended epoch
  that is skewed toward early times, a delay in reionization, or an
  early incomplete reionization epoch at high $z$ may increase this
  value \citep{Zahn12,Park13} to harmonize with the latest
  measurements of $\tau = 0.092 \pm 0.013$ as deduced by a joined
  analysis of CMB data by {\em Planck}, high-multipole experiments,
  WMAP polarization and baryon-acoustic oscillations
  \citep{Planck_cosmpar}.} with the cosmological constraints for
\hfesc$=0.20$ and $0.35$ ($\tau=0.055$ and $0.065$ respectively).

For the sake of completeness, we also consider a model summing up the
QSO contribution to the \hfesc$=0.2$ model (purple dot-dashed line,
\tfesc$=1$): including this contribution to the ionizing photon budget
induces a marked redshift evolution of the background\footnote{It is
  worth stressing that an evolution of the ionizing background can
  also be achieved by assuming a redshift dependent dust correction
  \citep[see e.g][]{Cucciati12}.}. The total background is compatible
with the \citet{BeckerBolton13} data, but it overpredicts previous
results at a 3 $\sigma$ level, thus favouring lower \hfesc~values. Of
course, different combinations of \hfesc~and \tfesc~lead to similar
predictions of the redshift evolution of the ionizing background.
Indeed, several theoretical works suggest an evolution of \tfesc~with
AGN luminosity \citep[see e.g.,][and references
  herein]{Giallongo12}. As an example, if we consider our reference
model for LBG contribution with \hfesc$=0.2$ and if we require the
total background to be compatible with the lowest redshift point in
\citet{BeckerBolton13}, we predict \tfesc$\sim0.25$ (assuming
\muvlim$=-23$) or \muvlim$\sim-26.5$ (assuming \tfesc$=1$). This makes
the case for a better estimate of \tfesc, in order to get an even
better constraint of \hfesc. This may be achieved by detailed
observation of $2 \lesssim z \lesssim 3$ sources.

\begin{figure}
  \centerline{ \includegraphics[width=9cm]{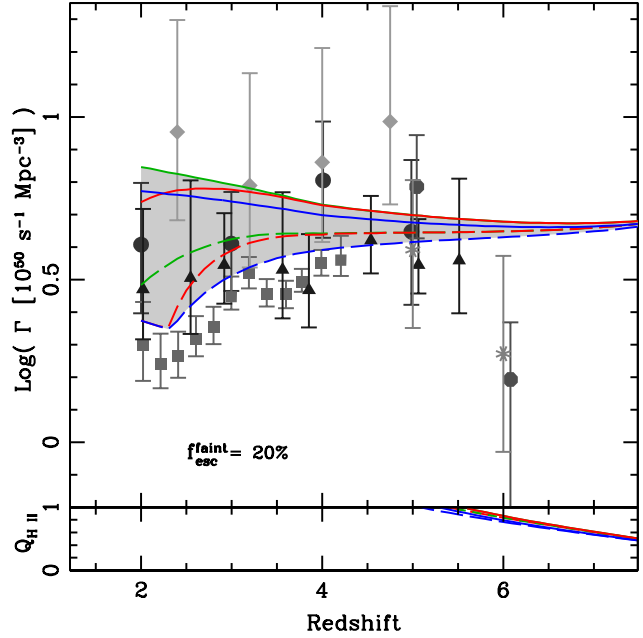} }
  \caption{Synthesis of the ionizing background. Datapoints and panels
    as in Fig.\ref{fig:fevo}. Solid (dashed) lines represent the LBGs
    contribution estimated from integration the LBG-LFs up to a
    magnitude limit defined by the characteristic (filtering) mass and
    assuming \fesc$=0.2$ for LBGs fainter than $M_{\rm break}=-18$
    (and \fesc$=0.05$ for brighter sources). Green, red and blue lines
    correspond to different thermal histories: pure photoheating,
    blazar heating and constant temperature ($T=10^4 {\rm K}$),
    respectively. Grey shading marks the allowed area between the
    different estimates of the filtering and the characteristic
    masses.}\label{fig:mainO}
\end{figure}
\begin{figure}
  \centerline{\includegraphics[width=9cm]{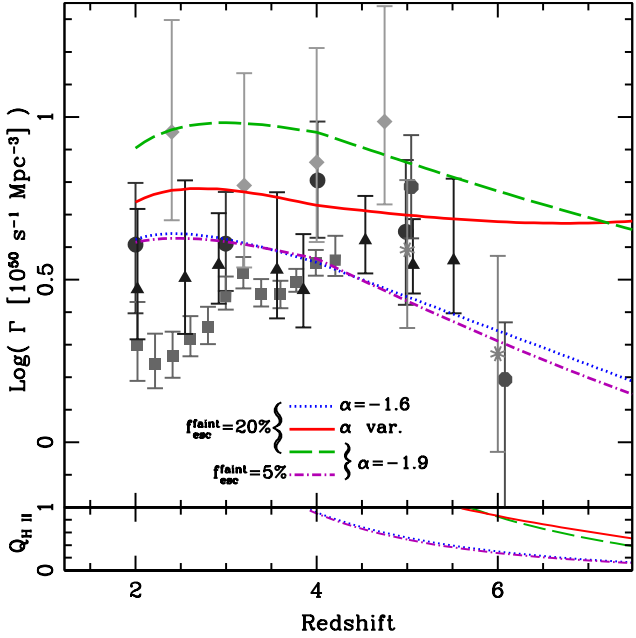} }
  \caption{Synthesis of the ionizing background. Datapoints, panels
    and solid red line as in Fig.\ref{fig:fevo}. Dotted and dashed
    lines represent the LBG contribution obtained by assuming
    \hfesc$=0.2$ and a redshift independent faint end slope for the
    UV-LF ($\alpha=-1.6$ and $-1.9$, respectively). The dot-dashed
    line refer to the LBG contribution obtained by assuming
    \hfesc$=0.05$ and $\alpha=-1.9$. For all lines we consider the
    blazar heating thermal history and characteristic mass
    case.}\label{fig:slope}
\end{figure}

We then explore the parameter space associated with our theoretical
model to check the robustness of our results. First of all, we test
the dependency of our results on the details of
equation~(\ref{eq:hfesc}): brightening the break magnitude $M_{\rm
  break}$ to $-17$ increases the predicted background by a few per
cents (leaving our conclusions unchanged), while decreasing the
fiducial value of \fesc~for bright galaxies has the opposite
effect. In Fig.~\ref{fig:mainO}, we then study the effect of the
uncertainties on \muvlim, by fixing \hfesc$=0.2$ and consider
different choices for the IGM thermal histories: green, red and blue
lines correspond to the predicted ionizing background for the pure
photoheating, blazar heating and constant temperature cases,
respectively, while solid (dashed) lines refer to characteristic
(filtering) masses. We mark the allowed area between the different
estimates of the filtering and the characteristic masses with grey
shading. The predicted background is almost constant with redshift for
models based on the characteristic mass, while it has a mild redshift
evolution for models based on the filtering mass. Within the
uncertainties all models agree with the lower observational estimates
for the background. It is worth stressing that the filtering mass
approach predicts that no galaxy fainter than our current accessible
depth should exist at $2 \lesssim z \lesssim 3$; this result is in
tension with recent findings of ultra-faint UV galaxies ($-19.5<M_{\rm
  UV}<-13$) at $z\sim2$ as magnified sources in the background of the
lower-redshift ($z\sim0.18$) lensing galaxy cluster Abell 1689
\citep{Alavi13}. On the other hand, at least for the characteristic
mass scenario, the mere existence of galaxies fainter than the limits
shown in Fig.~\ref{fig:m145lim} does not indicate a substantial
discrepancy between theory and observations. In fact, we assume
\muvlim~as an indication of the magnitude scale at which galaxy
formation becomes ineffective, i.e. the magnitude at which we expect
the UV-LF either to drop or to flatten, thus assuming a negligible
contribution to the background from fainter sources. From their
ultra-faint UV sample, \citet{Alavi13} conclude that the UV-LF is
consistent with a Schechter function until \muvlim$\sim -13$, but we
caution the reader that, at the faintest magnitudes, this result is
based on a limited sample with fairly relevant completeness
corrections. Our predictions thus make the search for faint galaxies
in this redshift interval, and the characterisation of their
properties, a relevant goal for future surveys.
\begin{figure}
  \centerline{\includegraphics[width=9cm]{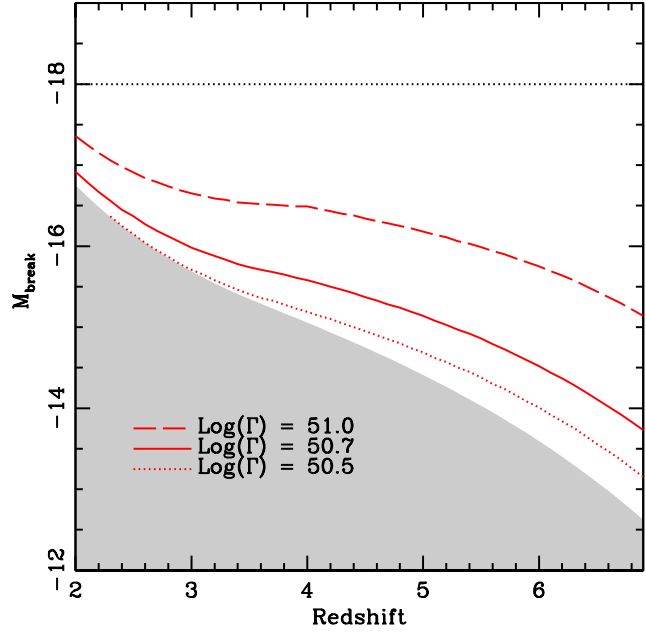} }
  \caption{Critical $M_{\rm break}$ required to reproduce a given
    background level with \hfesc$=1$. Dotted, solid and dashed line
    refer to three different background levels as noted in the figure
    legend. The horizontal dotted line represents the present limit
    for faint LBG searches \citep{Bouwens11a}. All lines refer to the
    blazar heating thermal history and characteristic mass
    scenario. Shaded area represents the region corresponding to
    magnitudes fainter than \muvlim.}\label{fig:mbreak}
\end{figure}

We then discuss the effect of the uncertainties on the faint-end slope
$\alpha$. In our standard prescription (red solid line -
\hfesc$=0.2$), we assume a redshift evolution for $\alpha$ at $z>4$,
following the results of \citet{Bouwens11a}; in Fig.~\ref{fig:slope}
we explore the effect of considering a redshift independent
$\alpha$. We consider two extreme values for $\alpha$ ($-1.6$ and
$-1.9$), which corresponds to the range of allowed values at $3<z<6$
\citep[see e.g., Fig.~2 in][]{Bouwens11a}. Our results clearly show
that the flat redshift evolution of the background is mainly due to
the assumed redshift evolution of $\alpha$ in the reference model,
while models with fixed slope predict an increasing $\Gamma$ with
decreasing redshift. The slope of this relation is shallower with
respect to the analogous relation in Fig.~\ref{fig:obs} (red lines -
where we use a fixed $L_{\rm lim}$ and a variable $\alpha$). The
dispersion of the models is of the same amplitude as in
Fig.~\ref{fig:fevo}. To test that our main conclusion still holds, we
add in Fig.~\ref{fig:slope} a model (purple dot-dashed line) with
$\alpha=-1.9$ and \hfesc$=0.05$: this realization clearly shows that
even in the most favourable case (steeper faint-end slope), a model
with \hfesc~as low as the current estimates at the bright-end, fails
to achieve cosmic reionization at $z\sim6$.

\section{Conclusions}\label{sec:concl}

In this paper, we study the nature of the sources responsible for
cosmic reionization by using the observed cosmic ionizing background
as a constraint. Our approach, under the hypothesis that cosmic
reionization is mainly driven by star forming galaxies, makes the case
for an increase of \fesc~in fainter sources. Overall, we thus propose
that a reliable measure of \fesc~for galaxies fainter than the present
observational limits ($M_{\rm break}=-18$) will provide valuable
insight into the galaxy population responsible for cosmic
reionization. Low values (\fesc$\lesssim0.1$) would imply either that
star forming galaxies are not the main contributors for cosmic
reionization (thus leaving room for more exotic explanations), or that
the bulk of the contribution comes from even fainter galaxies (with
even higher \fesc). Extrapolating our approach, it is then possible to
give a rough estimate of the critical luminosity, beyond which low
values of \fesc would imply that star forming galaxies {\it are not}
the main contributors to the background. We define it as the $M_{\rm
  break}$ value required to reproduce a given background level with
\hfesc$=1$. We show the predicted $M_{\rm break}(z)$ for three
different background levels and our reference model in
Fig.~\ref{fig:mbreak}. On the other hand, high values of \fesc
($\sim0.35$) would favour the highest observational estimates of
$\Gamma$, i.e. lower IGM temperatures, thus providing additional
insight into the IGM properties at the relevant redshifts (conversely,
a better agreement on the observational determinations of $\Gamma$
will allow tighter constraints on the \fesc). Even higher values
(\fesc$\gtrsim0.4$), on the other hand, lead to an overpredictions of
the ionizing background at higher redshift, thus implying a large
object-to-object scatter in \fesc. Our predictions can be tested
directly in the context of present and future facilities (e.g., Hubble
Space Telescope/Wide Field Camera 3, James Webb Space Telescope,
Extremely Large Telescope) or through indirect techniques \citep[see
  e.g.,][]{Robertson10, Vanzella12b} and show the importance of a
clean determination of \fesc~for the fainter galaxies accessible, in
order to get a better insight into the nature of the sources
responsible for cosmic reionization.

\section*{Acknowledgements}

We thank the anonymous referee for comments that helped to improve the
paper. We also thank Volker Springel and Michele Trenti for
enlightening discussions and suggestions. FF and CP acknowledge
financial support from the Klaus Tschira Foundation. FF acknowledges
financial support from the Deutsche Forschungsgemeinschaft through
Transregio 33, ``The Dark Universe''. FF, SC and EV acknowledge
financial contribution from the grant PRIN MIUR 2009 ``The
Intergalactic Medium as a probe of the growth of cosmic structures''.

\bibliographystyle{mn2e}
\bibliography{fontanot}

\end{document}